# Titanium-based kagome superconductor CsTi$_3$Bi$_5$ and topological states

Haitao Yang[1,2,3#], Zhen Zhao[1,2#], Xin-Wei Yi[2,3#], Jiali Liu[1,2#], Jing-Yang You[2,3], Yuhang Zhang[1,2], Hui Guo[1,2,3], Xiao Lin[2,3], Chengmin Shen[1,2], Hui Chen[1,2,3], Xiaoli Dong[1,2,3*], Gang Su[2,3*], and Hong-Jun Gao[1,2,3*]

[1] Beijing National Center for Condensed Matter Physics and Institute of Physics, Chinese Academy of Sciences, Beijing 100190, PR China

[2] School of Physical Sciences, University of Chinese Academy of Sciences, Beijing 100190, PR China

[3] CAS Center for Excellence in Topological Quantum Computation, University of Chinese Academy of Sciences, Beijing 100190, PR China

[#]These authors contributed equally to this work

[*]Correspondence to: hjgao@iphy.ac.cn, gsu@ucas.ac.cn, dong@iphy.ac.cn

Since the discovery of a new family of vanadium-based kagome superconductor $AV_3Sb_5$ (A=K, Rb, and Cs) with topological band structures, extensive effort has been devoted to exploring the origin of superconducting states and the intertwined orders [1-13]. Meanwhile, searching for new types of superconductors with novel physical properties and higher superconducting transition temperatures has always been a major thread in the history of superconductor research. Here we report a successful fabrication and the topological states of a Titanium-based kagome metal $CsTi_3Bi_5$ ($CT_3B_5$) crystal. The as-grown $CT_3B_5$ crystal is of high quality and possesses a perfect two-dimensional kagome net of Titanium. The superconductivity of the $CT_3B_5$ crystal shows that the critical temperature $T_c$ is of ~4.8 K. First-principle calculations predict that the $CT_3B_5$ has robust topological surface states, implying that $CT_3B_5$ is a $Z_2$ topological kagome superconductor. This finding provides a new type of superconductors and the base for exploring the origin of superconductivity and topological states in kagome superconductors.

We first present the crystal structural geometry of CsTi₃Bi₅ (CT₃B₅). The crystalline structure of the CT₃B₅ is of hexagonal with the space group of P6/mmm (Fig. 1**a**) [1]. The compound forms a layered structure. The perfect kagome net of Ti atoms mixed with a simple triangular net of Bi1 atoms is located in the middle layer, which is sandwiched by two additional honeycomb layers of Bi2 atoms (Fig. 1**b**). The upper and lower triangular layers of Bi metal atoms have large bond distances with respect to the middle Ti layer and are loosely bonded to them.

We synthesize the CT₃B₅ crystal through a self-flux method. A typical CT₃B₅ crystal with a lateral size of over 3 mm and regular hexagonal morphology is shown in the inset of Fig. 1**c**. The representative x-ray diffraction (XRD) pattern of the CT₃B₅ crystal confirms the pure phase of the as-grown single crystal with a preferred [001] orientation (Fig. 1**c**). The rocking curve obtained from the (004) reflection shows a full-width-at-half-maxima (FWHM) of ~0.078°, demonstrating the single crystal nature of the as-grown CT₃B₅ (Fig. 1**d**). The peaks in the XRD curve agree with the theoretical prediction (Fig. S1), indicating that the atomic structure of the as-grown CT₃B₅ is consist with the one shown in Fig. 1**a**. The lattice parameters $a$, $b$, and $c$ are measured to be 5.839, 5.839 and 9.295 Å, respectively, by single crystal diffraction (Fig. S2) [7]. The layered structures are confirmed by the scanning electron microscopy (SEM) measurements (Fig. S3). The characterizations above, therefore, demonstrate the high-quality of the as-synthesized CT₃B₅.

We then verify the superconductivity by collecting the zero-field cooled (ZFC, hollow symbols) and field cooled (FC, solid symbols) magnetic susceptibility (Fig. 2**a**). The onset superconducting transition temperature is 4.8 K. The existence of superconducting phase was unambiguously conformed by the Meissner effect. A superconducting volume fraction above 50% under a magnetic field of 0.05 mT was obtained at 1.8 K, indicating that the superconductivity is bulk in nature. The heavily field-dependent diamagnetism shown in Fig.2**a** implies an unusual superconducting feature of this new type superconductor. The measurements of lower critical fields $Hc_1(T)$ can also provide important information on superconductivity in the as-grown kagome metal CT₃B₅, which are subtracted from isothermal magnetization $M(H)$ with magnetic field along the $c$ axis (Fig. 2**b**) and $ab$ plane (Fig. 2**c**), respectively. The $H_{C1}$ data follow a single-band s-wave relation with the fitted values $\Delta(0) = 1.76\ k_B T_c$, and $\mu_0 H_{C1}^c(0)$ = 13.3±0.3 mT, $\mu_0 H_{C1}^{ab}(0)$ = 19.0 ±0.3 mT , respectively (Fig. 2**d**), indicating a conventional BCS character. Moreover, the abnormal $H_{C1}$ anisotropy of CT$_3$B$_5$ ($H_{c1}^{ab} > H_{c1}^{c}$), resembling the CsV$_3$Sb$_5$ case [14], implies a similarity in Fermi surface topology between these two types of kagome superconductors

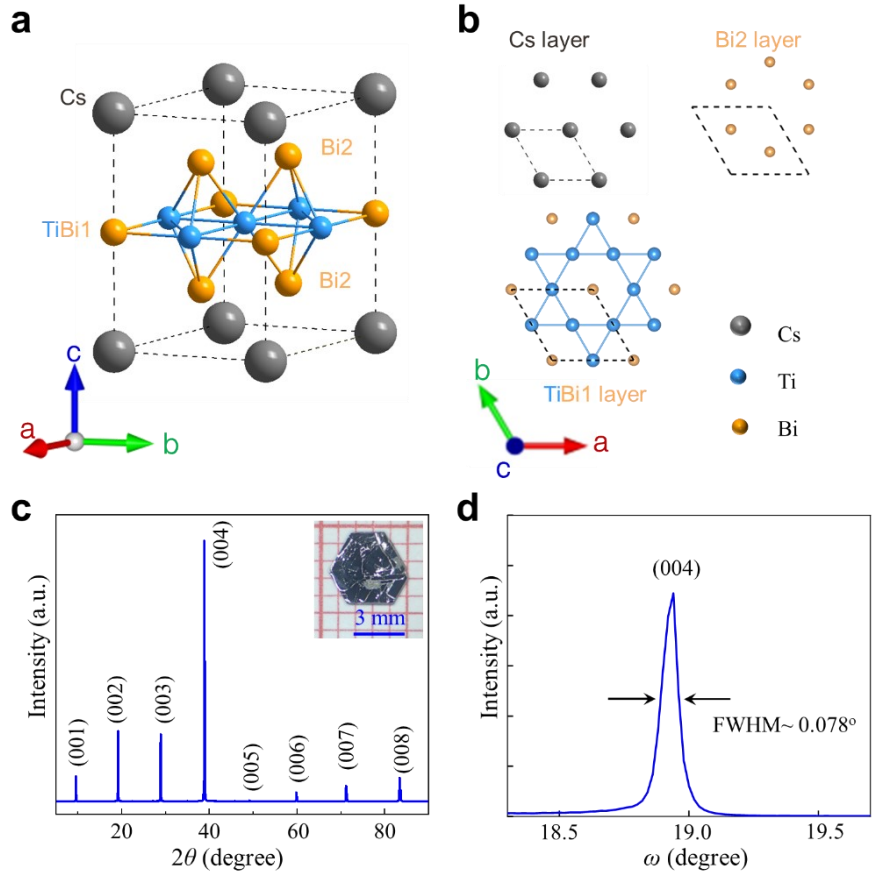

**Fig. 1| Atomic structure of the Ti-based kagome superconductor CsTi$_3$Bi$_5$. a**. Crystal structure for CT$_3$B$_5$ with Cs atoms in black, Bi atoms in light orange, Ti atoms in azure. **b**, Atomic arrangements of the Cs, TiBi and Bi atomic layers. The dashed lines represent a unit cell. Cs layer forms a hexagonal lattice, while the Bi2 layer consists of a honeycomb lattice. The perfect kagome net of Ti atoms is mixed with a simple triangular net of Bi1 atoms. **c,** XRD pattern of the CT$_3$B$_5$ single crystal, revealing that the crystal surface is parallel to the (001) plane. Inset: a photo of a typical CT$_3$B$_5$ single crystal. **d**, The x-ray rocking curve of the (004) reflection shows a small FWHM of 0.078°.

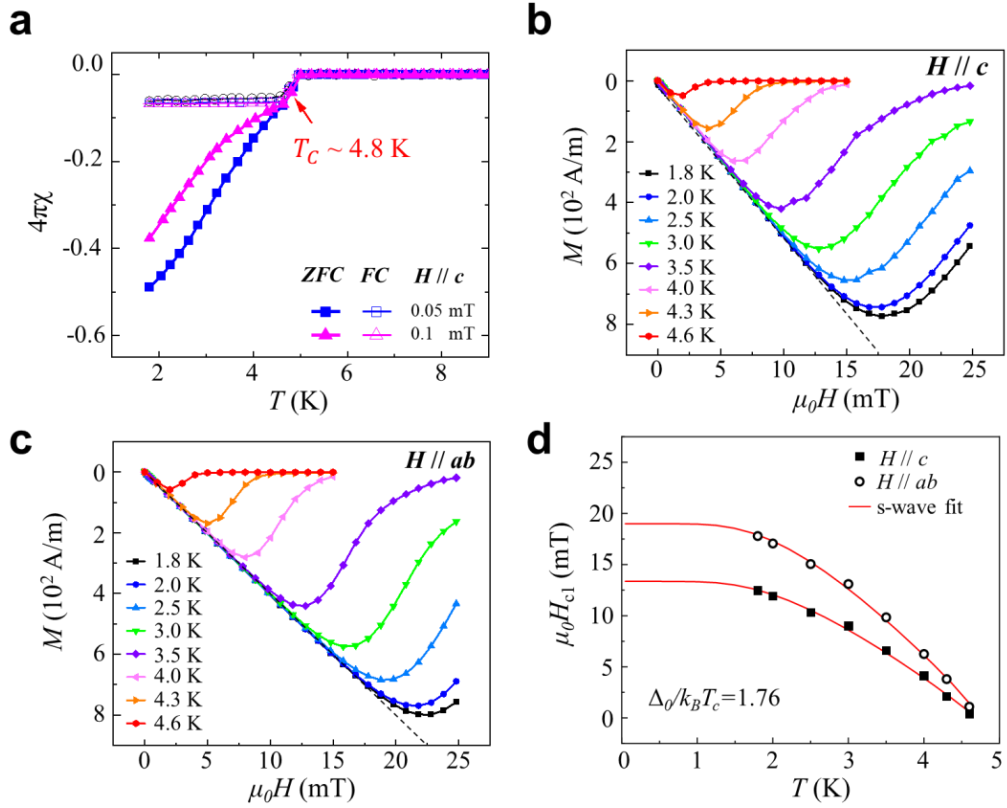

**Fig. 2| Magnetization measurements of the CsTi$_3$Bi$_5$ single crystal. a,** The magnetic susceptibilities, corrected for the demagnetization factor, at different field for *H // c* with ZFC and FC progress between 1.8 K and 10 K. **b, c,** The isothermal magnetization at various temperatures with magnetic fields along the *c* axis (b) and *ab* plane (c). **d,** Temperature dependence of lower critical fields $\mu_0 H_{c1}$ for *H // c* and *H // ab*, respectively.

Superconductivity in CT$_3$B$_5$ is also confirmed by electrical transport measurements. The CT$_3$B$_5$ sample shows a metallic behavior (Fig. 3**a**) with a superconducting transition at onset $T_c \sim$ 4.8 K and zero-resistivity at ~ 3.6 K (Fig. 3**b**). It is noticeable that this new kagome superconductor CT$_3$B$_5$ exhibits *T*-linear resistivity over a wide temperature range (~70 – 300 K), which might be caused by electron-phonon scatterings. The residual resistance ratio (RRR) subtracted from the temperature dependence of the resistivity is about 26, further supporting the relatively high quality of the as-grown CT$_3$B$_5$ crystals.

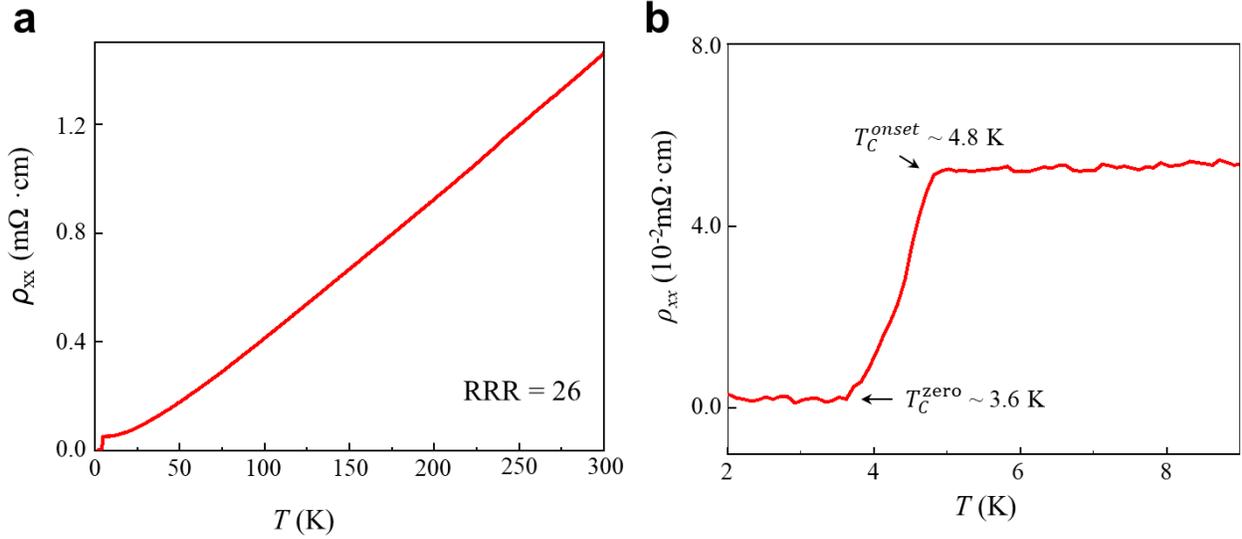

**Fig. 3| Temperature dependence of resistivity of the CsTi$_3$Bi$_5$ single crystal. a,** Temperature-dependent resistance between 2 K and 300 K under zero magnetic field, showing the RRR is about 26. **b,** Temperature-dependent resistivity between 2 K and 9 K under zero magnetic field, showing the $T_C^{onset}$=4.8 K and $T_C^{zero}$ ~ 3.6 K.

Finally, we predict the physical property of the CT$_3$B$_5$ crystal with first-principles DFT calculations. The electronic band structure and corresponding density of states (DOS) with spin-orbit coupling (SOC) for CT$_3$B$_5$ (Fig. 4**a**) show that the band structures in k$_z$=0 and k$_z$=π planes are very similar, indicating a strong 2D characteristic similar to AV$_3$Sb$_5$. Ti and Bi atoms mainly contribute to the DOS near the Fermi level, forming a small peak, and at slightly above and below the Fermi level there are dips of the DOS owing to the energy gap opened by the strong SOC. We note that the energy bands colored by red, blue and green in Fig. 4**a** give rise to a strong topological Z$_2$ index, resulting in topologically nontrivial surface states near the Fermi level (Figs. 4**b** and **c**), which indicates that CT$_3$B$_5$ is a Z$_2$ topological superconductor. Combining the time-reversal and inversion symmetries in CT$_3$B$_5$, we obtain Z$_2$ topological invariant (Fig. 4**d**) of several bands near Fermi level by calculating the parity of the wave functions at all time-reversal invariant momenta (TRIM)[15]. Thus, the CT$_3$B$_5$ reveals the coexistence of superconductivity and nontrivial topological surface states.

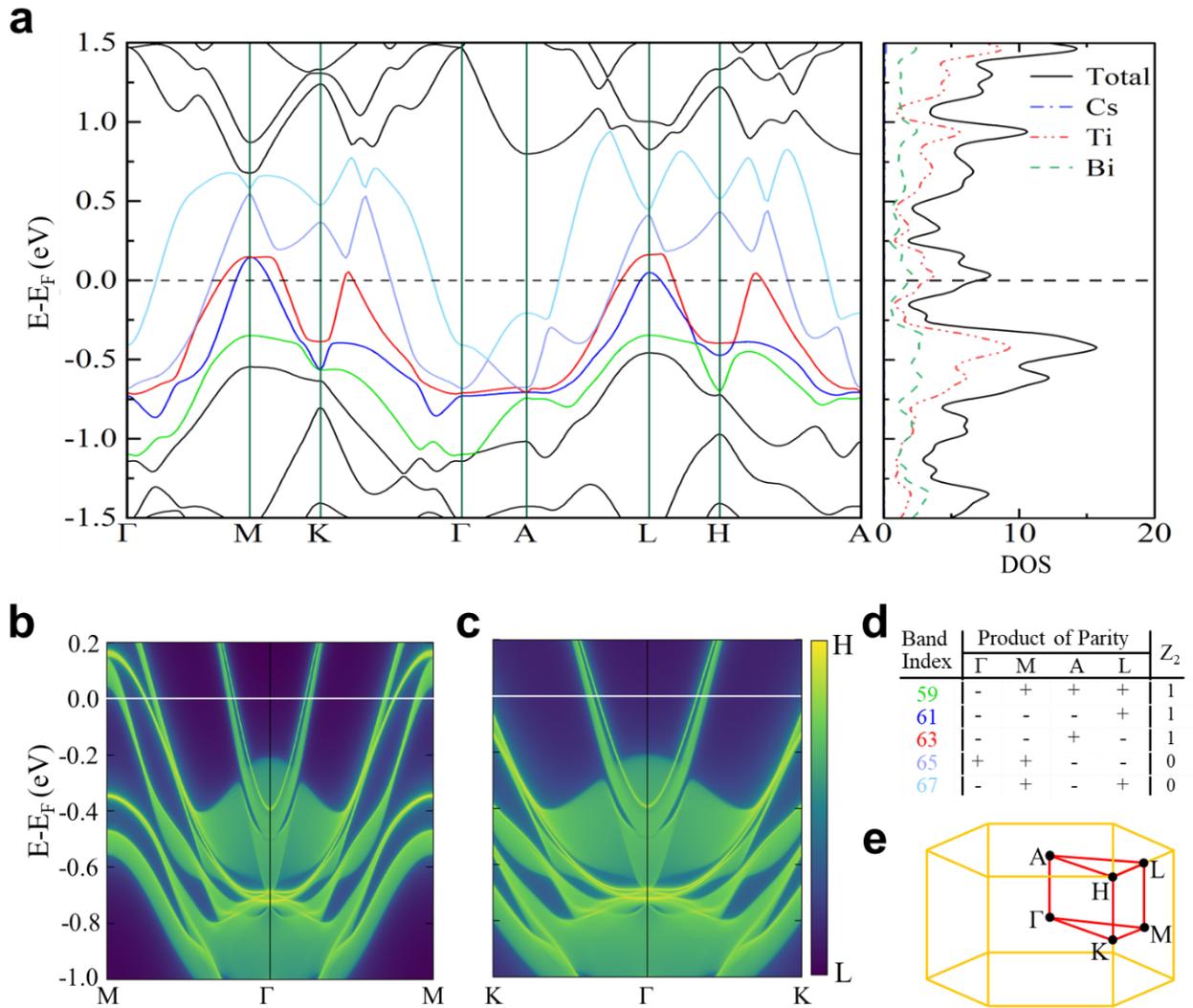

**Fig. 4| Calculated band structures and $Z_2$ band indices of CsTi$_3$Bi$_5$ single crystal. a,** The electronic energy bands and density of states calculated with spin-orbit coupling for CT$_3$B$_5$. Energy bands near Fermi level are drawn in different colors. **b,c,** The surface states along (**b**) M-Γ-M and (**c**) K-Γ-K paths projected on (001) plane for CT$_3$B$_5$. **d,** Product of parity of four time reversal invariant momenta and $Z_2$ indices of the bands near Fermi level. The colors of band indices are consistent with (**a**). **e,** The Brillouin zone with high symmetry paths.

In summary, we have successfully fabricated the Ti-based kagome superconductor CsTi$_3$Bi$_5$ crystal of high quality and discovered the bulk superconductivity for the first time. The transition superconducting temperature is of 4.8 K. The DFT calculations show that the CT$_3$B$_5$ superconductor has robust topological surface states, implying that CT$_3$B$_5$ is a Z$_2$ topological kagome superconductor and Majorana zero modes are expected to be observed at the CT$_3$B$_5$ surface. These findings provide a new platform for future studies on topological superconductivity, correlated electronic states, and topological quantum computations.